\documentclass[aps,prc,showpacs,amsmath,amssymb]{revtex4}

\usepackage{epsfig}

\begin{document}


\title{Au+Au  Collisions: the  Suppression  of High  Transverse
  Momentum in $\pi^0$ Spectra.}

\author{D.~E.~Kahana}
\affiliation{31 Pembrook Dr.,
Stony Brook, NY 11790, USA}
\author{S.~H.~Kahana}
\affiliation{Physics Department, Brookhaven National Laboratory\\
   Upton, NY 11973, USA}

\date{\today}  
 
\begin{abstract}

Au+Au, $\sqrt{s} = 200$ A GeV measurements at RHIC, obtained with
the PHENIX, STAR, PHOBOS and BRAHMS detectors, have all indicated
a  suppression  of   production,  relative  to  an  appropriately
normalized  NN  level.   For  central  collisions  and  vanishing
pseudo-rapidity   these  experiments   indicate   a  considerable
reduction, relative to that nominally expected from equivalent PP
measurements, in  charged and particle  production, especially at
mid-  to  large transverse  momenta.   In  the PHENIX  experiment
similar  behavior has been  reported for  $\pi^0$ spectra.   In a
recent  work~\cite{luc4brahms}  on  the presumably  simpler  D+Au
interaction, to be considered perhaps  as a tune-up for Au+Au, we
reported on  a hadronic cascade  mechanism which can  explain the
observed   moderatereduced   $p_\perp$   suppression  at   higher
pseudorapidity  as  well  as   the  Cronin  enhancement  at  mid-
rapidity. Here we present the  extension of this work to the more
massive ion-ion collisions.

\end{abstract}


\maketitle 
\section{Introduction}

The specific question at hand  in this work is the suppression of
medium  to  high  transverse  momentum  yields  observed  in  the
experimental   measurements~\cite{brahms,phenix,phobos,star}  for
Au+Au  at 130 GeV and  200 GeV.   The experimental  results have
focused on the $\eta$ and $p_\perp$-dependence of the ratio

\begin{equation}
\text{R}[\text{AA/NN}] =
\left(\frac{1}{\text{N}_{coll}}\right)
\frac{
[d^2N^{ch}/dp_\perp\,d\eta]\,\,({\text{AA}})
}
{
[d^2N^{ch}/dp_\perp\,d\eta]\,\,({\text{NN}})
}
,
\end{equation}

\noindent  where  $\text{N}_{coll}$  is  a calculated  number  of
binary NN collisions occurring in Au+Au at a designated energy and
centrality. One can  also, of course, just copare  directly to the
data without reference to ratios.

The simulation code LUCIFER,  developed for high energy heavy-ion
collisions  has  previously been  applied  to  both SPS  energies
$\sqrt{s}=(17.2,20)$  A GeV~\cite{lucifer1,lucifer2,luc3}  and to
RHIC       energies      $\sqrt{s}=(56,130,200)$       A      GeV
~\cite{lucifer1,luc3}.   Although nominally intended  for dealing
with  soft, low $p_T$,  interaction it  is possible  to introduce
high  $p_T$ hadron  spectra via  the  NN inputs,  which form  the
building  blocks of  the  simulations, and  to  then examine  the
effect of  rescattering, and  concommittant energy loss,  on such
spectra~\cite{luc4brahms}.The  simulation  is  divided  into  two
phase  I  and II,  with  all  the  rescattering and  energy  loss
restricted to  the second and considerably  reduced energy stage.
The  first  stage sets  up  the  participants,  both mesonic  and
baryonic, their  four momenta and positions  for the commencement
of  the  cascade  in  II.   The  second  stage  energy  loss  and
interactions  within a  hadronic fluid  play  a key  role in  the
eventual suppression of the transverse momentum distribution.

For completeness  we present a brief description  of the dynamics
of  this Monte  Carlo  simulation~\cite{luc4brahms}.  Many  other
simulations  of   heavy  ion  collisions  exist   and  these  are
frequently  hybrid in  nature,  using say  string  models in  the
initial
state~\cite{rqmd,rqmd2,bass1,frithjof,capella,werner,ko,ranft} 
together  with final state hadronic  collisions, while some
codes           are           purely          or           partly
partonic~\cite{boal,eskola,wang,wang2,geiger,bass2}   in  nature.
Our  approach  is   closest  in  spirit  to  that   of  RQMD  and
K.~Gallmeister, C.~Greiner,  and Z.~Xu~\cite{greiner} as  well as
work  by  W.~Cassing~\cite{rqmd,greiner,cassing}.  Certainly  our
results seem to parallel those  of the latter authors.  Both seek
to  separate  initial  perhaps  parton dominated  processes  from
hadronic  interactions  occurring at  some  intermediate but  not
necessarily late time.

The  purpose of  describing such  high energy  collisions without
introducing the  evident parton nature  of hadrons, at  least for
soft  processes,  was  to  set  a baseline  for  judging  whether
deviations  from the simulation  measured in  experiments existed
and  could  then  signal  interesting  phenomena.   The  division
between soft  and hard processes,  the latter being  in principle
described  by  perturbative  QCD,  is  not  necessarily  easy  to
identify in  heavy ion data,  although many authors  believe they
have    accomplished     this    within    a    gluon-saturation
configuration~\cite{saturation,cgc1,cgc2}.   For  both D+Au  and
Au+Au systems we separate the  effects of a second stage, a lower
energy  hadronic  cascade,  from  those  of the  first  stage,  a
parallel   rather   than    sequential   treatment   of   initial
(target)-(projectile) NN interactions.

In the  present work, even absent some  energy-loss effects which
one  might   anticipate  in  the   initial  phase,  we   do  find
considerable  suppression   of  the  Au+Au   transverse  momentum
spectrum at cental  rapidity. One might say that  the first stage
of our  simulation, involving  the collective interaction  of the
initially present nucleons,  produces a ``hot-gas'' of prehadrons
which  are  considerably  cooled  in an  inevitable  final  state
cascade.    This  cooling   constitutes   the  observed   ``jet''
suppression, not  surprisingly a suppression  appreciably greater
for  Au+Au than  for  D+Au~\cite{brahms,luc4brahms}.  The  second
stage  II,  a  true  cascade,  critically  includes  energy  loss
effects.  The cascade  interactions inevitably soften the initial
$p_\perp$  distribution,  as  II  involves  considerably  reduced
energy processes.

\section{The Simulation}
\subsection{Stage I}

The first  stage I of LUCIFER considers  the initial interactions
between the separate  nucleons in the colliding ions  A+B, but is
not a cascade.  The  totality of events involving each projectile
particle  happen  essentially  together   or  one  might  say  in
parallel.   Neither  energy   loss  nor  creation  of  transverse
momentum  ($p_\perp$)  are  permitted  in  stage  I,  clearly  an
approximation.           A          model          of          NN
collisions~\cite{lucifer1,lucifer2},   incorporating  most  known
inclusive cross-section and multiplicity data, guides stage I and
sets up the initial conditions for stage II.  The two body model,
clearly an  input to  our simulation, is  fitted to  the elastic,
single   diffractive  (SD)   and  non-single   diffractive  (NSD)
aspects~\cite{goulianos}   of   high   energy   $PP$   collisions
~\cite{ua5,ua1}  and  $P\bar   P$  data~\cite{fermilab}.   It  is
precisely  the  energy   dependence  of  the  cross-sections  and
multiplicities of  the NN input  that led to our  successful {\it
prediction}~\cite{lucifer2} of the rather small ($13\%$) increase
in  $dN^{ch}/d\eta$ between  $\sqrt{s}=130$ and  $\sqrt{s}=200$ A
GeV, seen in the PHOBOS data~\cite{phobos1}.

We  find that  the  addition  to stage  I  baryons of  transverse
momentum by  collision dependent random  walk produces in  fact a
rather  `hot' hadronic  medium,  in effect  a strong  A-dependent
Cronin~\cite{cronin}  effect,  a  medium  which  is  subsequently
cooled by the final state  cascading collisions in stage II.  The
comparison  of the initial  and final  $p_T$ spectra  provides an
alternative   measure   of  suppression   to   the  above   ratio
{R}[{AA/NN].

A history of  the collisions that occur between  nucleons as they
move along straight  lines in stage I is  recorded and later used
to  guide the  determination of  multiplicity.   Collision driven
random walk fixes the $p_\perp$  to be ascribed to the baryons at
the  start of stage  II.  The  overall multiplicity,  however, is
subject  to  a  modification,  based  as we  believe  on  natural
physical requirements~\cite{lucifer1}.

If a sufficiently hard  process, for example Drell-Yan production
of  a lepton  pair at  large mass  occurred, it  would lead  to a
prompt  energy loss  in stage  I.  Hard  quarks and  gluons could
similarly be  entered into the particle lists  and their parallel
progression followed.  This has not yet been done.  One viewpoint
and justification for our approach for hadrons could be to say we
attempt to  ignore the direct  effect of colour on  the dynamics,
projecting  out  all states  of  the  combined system  possessing
colour.  In  such a situation  there should be a  duality between
quark-gluon or hadronic treatments.
 
The  collective/parallel method  of treating  many  NN collisions
between the target and projectile is achieved by defining a group
structure for  interacting baryons.  This is  best illustrated by
considering  a prototypical  proton-nucleus  (P+A) collision.   A
group is defined by spatial contiguity.  A nucleon at some impact
parameter  $b(\bar{x}_\perp)$  is  imagined  to  collide  with  a
corresponding  `row'  of   nucleons  sufficiently  close  in  the
transverse  direction to the  straight line  path of  the proton,
{\it   i.~e.}~within   a  distance   corresponding   to  the   NN
cross-section.   In   a  nucleus-nucleus  (A+B)   collision  this
procedure is generalized by making  two passes: on the first pass
one includes all  nucleons from the target which  come within the
given  transverse distance  of some  initial  projectile nucleon,
then on the  second pass one includes for  each target nucleon so
chosen, all of those  nucleons from the projectile approaching it
within the  same transverse distance.  This  totality of mutually
colliding   nucleons,   at   more   or  less   equal   transverse
displacements,  constitute  a  group.  The  procedure  partitions
target and projectile nucleons into a set of disjoint interacting
groups as well as a set of non-interacting spectators in a manner
depending on the overall  geometry of the A+B collision.  Clearly
the largest groups in P+A will,  in this way, be formed for small
impact parameters  $b$; while for the  most peripheral collisions
the groups  will almost always  consist of only one  colliding NN
pair. Similar conclusions hold in the case of A+B collisions.
     
In stage II of the  cascade we treat the entities which rescatter
as  prehadrons.  These  prehadrons, both  baryonic or  mesonic in
type, are not the  physical hadron resonances or stable particles
appearing in  the particle  data tables, which  materialize after
hadronisation.   Importantly prehadrons  are allowed  to interact
starting   at    early   times,   after    a   short   production
time~\cite{boris}, nominally  the target-projectile crossing time
$T_{AB}  \sim  2R_{AB}/\gamma$.  In  practice  we  find that  the
effective  time  delay for  prehadronic  collisions (stage  II)is
appreciable some 0.2 t0 0.4 fm/c in the center of momentum frame.
The mesonic  prehadrons are imagined  to have ($q \bar  q$) quark
content   and  their   interactions  are   akin  to   the  dipole
interactions included in models  relying more closely on explicit
QCD~\cite{boris,mueller},  but  are  treated here  as  colourless
objects.

Some  theoretical  evidence   for  the  existence  of  comparable
colourless structures is  given by Shuryak and Zahed~\cite{zahed}
and  by certain lattice  gauge studies~\cite{lattice}.   In these
latter  works  a basis  is  established  for  the persistence  of
loosely  bound  or  resonant   hadrons  above  the  QCD  critical
temperature $T_c$, indeed to  $T \sim (1.5-2.0)\times T_c$.  This
implies a persistence to  much higher transverse energy densities
$\rho(E) \sim (1.5-2.0)^4 \rho_c$, hence to the early stages of a
RHIC collision.  Accordingly, we  have incorporated into stage II
{\it hadron  sized} cross-sections for the  interactions of these
prehadrons,  although early  on it  may in  fact be  difficult to
distinguish  their colour  content.   Such larger  cross-sections
indeed  appear  to  be  necessary  for  the  explanation  of  the
apparently  large  elliptical  flow  parameter $\nu_2$  found  in
measurements~\cite {molnar,flow}.

The  prehadrons, which when  mesonic may  consist of  a spatially
close, loosely correlated quark  and anti-quark pair, are given a
mass spectrum between $m_\pi$ and $1.2$ GeV, with correspondingly
higher upper  and lower  limits allowed for  prehadrons including
strange  quarks.  The  Monte-Carlo  selection of  masses is  then
governed by a Gaussian distribution,
\begin{equation}
P(m)= \exp(-(m-m_0)^2/w^2),
\end{equation}
with $m_0$ a selected  center for the prehadron mass distribution
and $w=m_0/4$ the width.  For non-strange mesonic prehadrons $m_0
\sim 800$ MeV,  and for strange $m_0\sim 950$  MeV. Small changes
in $m_0$ and $w$ have little effect since the code is constrained
to fit hadron-hadron, data.

Too high an  upper limit for $m_0$ would  destroy the soft nature
expected for most prehadron  interactions when they finally decay
into `stable' mesons.   The same proviso can be  put in place for
prebaryons, restricting  these to a  mass spectrum from  $m_N$ to
$2$ GeV.  However, in  the present calculation the prebaryons are
for simplicity taken just to  be the normal baryons.  The mesonic
prehadrons  have  isospin   structure  corresponding  to  $\rho$,
$\omega$, or $K^*$, while the  baryons range across the octet and
decuplet.

Creating  these intermediate  degrees of  freedom at  the  end of
stage I  simply allows the original nucleons  to distribute their
initial energy-momentum  across a larger basis of  states or Fock
space, just  as is done in  string models, or for  that matter in
partonic   cascade   models.    Eventually,  of   course,   these
intermediate  objects decay  into physical  hadrons and  for that
purpose we assign a  uniform decay width $\sim \Gamma_{f}$, which
then  plays the  role of  a  hadronisation or  formation time,  $
1/\Gamma_{f} \sim 1 fm/c$.

\subsection{Elementary Hadron-Hadron Model}

The underlying  NN interaction structure  involved in I  has been
introduced  in  a  fashion  dictated  by  standard  proton-proton
modeling~\cite{goulianos}.   A  division  is made  into  elastic,
single    diffractive    (SD)    and    non-single    diffractive
components. Fits  are obtained  to the existing  two-nucleon data
over  a  broad  range  of  energies  {sqrt(s)},  using  the  same
prehadrons introduced above. No rescattering, only decay of these
intermediate   structures   is  permitted   in   the  purely   NN
calculation.   Specifically   the  meson-meson  interactions  are
scaled  to $4/9$  of the  known  NN cross-sections,  thus no  new
parameters  are  invoked.  Indeed,  since  only  known data  then
constrains  the  prehadronic  interaction,  this  approach  is  a
parameter-free input to the AA dynamics.

\subsection{Groups}

Energy  loss  and  multiplicity  in  each group  of  nucleons  is
estimated from  the straight  line collision history  recorded in
stage  I.   To  repeat,  transverse  momentum  of  prebaryons  is
assigned by a  random walk having a number of  steps equal to the
number  of  collisions  suffered.   The multiplicity  of  mesonic
prehadrons cannot be similarly directly estimated from the number
of NN collisions in a group.  We argue~\cite{gottfried} that only
spatial densities  of generic prehadrons~\cite{lucifer1,lucifer2}
below some maximum are allowable, {\it viz.}  the prehadrons must
not  overlap  spatially at  the  beginning  of  stage II  of  the
cascade.  The prehadrons then constitute an incompressible fluid.

The  KNO scaled  multiplicity  distributions, present  in our  NN
modeling  are  sufficiently  long-tailed  that  imposing  such  a
restriction on overall multiplicity  can for larger nuclei affect
results   even    in   P+A   or   D+A    systems.    In   earlier
work~\cite{lucifer1,luc3} the centrality dependence of $dN/d\eta$
distributions for RHIC energy Au+Au collisions was well described
with such a  density limitation on the prehadrons,  which was not
carried out  as efficaciously as in the  present work, especially
with to respect to highly peripheral collision.

Again, importantly, the  cross-sections in prehadronic collisions
were   assumed  to   be   the  same   size   as  hadronic,   {\it
e.~g.}~meson-baryon  or meson-meson  values {\it  etc.}~,  at the
same center  of mass energy, thus introducing  no additional free
parameters into  the model.   Where the latter  cross-sections or
their  energy  dependences  are  inadequately known  we  employed
straightforward quark  counting to  estimate the scale.   In both
SPS,  Pb+Pb and  RHIC Au+Au  events  at several  energies it  was
sufficient in earlier works to impose this constraint at a single
energy.  In  the present  modelling  the  densities are  actually
estimated  internally and  thereby  restricted dynamically.   The
inherent  energy dependence in  the KNO-scaled  multiplicities of
the NN inputs and the geometry then take over.

\subsection{High Transverse Momenta}

One  question which  has yet  to be  addressed concerns  the high
$p_\perp$  tails  included in  our  calculations.  In  principle,
LUCIFER  is  applicable  to  soft processes  {\it  i.~e.}~at  low
transverse momentum.  Where the cutoff in $p_\perp$ occurs is not
readily apparent. In any case we can include high $p_\perp$ meson
events through  inclusion in the  basic hadron-hadron interaction
which  is  of  course  an  input  rather than  a  result  of  our
simulation.   Thus   in  Fig(1)  we  display   the  NSD  $(1/2\pi
p_\perp)(d^2N^{charged}/dp_\perp\,d\eta)$   from  UA1~\cite{ua1}.
One can use  a single exponential together with  a power-law tail
in $p_\perp$, or alternatively two exponentials, to achieve a fit
of the  output in PP to  UA1 $\sqrt{s}$=200 GeV  data. A sampling
function of the form
\begin{equation}
f = p_\perp (a \exp(-p_\perp/w) + b / ((1 + (r / \alpha)^ \beta)),    
\end{equation}
gives a  satisfactory fit to the PP data in the Monte-Carlo.

Additionally, since we,for  the moment, constrain our comparisons
to the production  of neutral pions in Au+Au  we also present, in
Fig.(2)  the PHENIX~\cite{phenixpp}  midrapidity  $p_\perp$ yield
for NN together with  our representation of this spectrum.  These
NN  generated $p_\perp$  spectra,  inserted into  the code,  were
first   applied   to   the   meson  $p_\perp$   distribution   in
D+Au~\cite{luc4brahms} and now of course to Au+Au.  No correction
is  made for  possible  energy  loss in  stage  I, an  assumption
parallel  to  that  made  by   the  BRAHMS  and  all  other  RHIC
experiments,  in analysing  $p_\perp$ spectra  and multiplicities
irrespective  of  low or  high  values.   However, some  explicit
energy loss is definitely present  in the collisions of stage II,
in the dynamics through energy conservation and by a modification
of the $p_\perp$ spectra with energy.

Since  we impose  energy-momentum conservation  in each  group, a
high $p_\perp$  particle having say, several  GeV/c of transverse
momentum,  must   be  accompanied  in   the  opposite  transverse
direction   by  one   or  several   compensating   mesons.   Such
high-$p_\perp$  leading particles  are not  exactly jets,  to the
extent that  they did not  originate in our simulation  from hard
parton-parton  collisions,  but  they  yield  much  of  the  same
observable experimental behaviour. It must be emphasized that the
totality  of $p_\perp$  events  is small,  certainly  for the  NN
collisions seen in Figs.(1,2)) and also as we will see for any AA
events.  In fact  some $90\%$ of all final  mesonic yields occurs
for $p_\perp \le  0.7$.  This implies that our  treatment of such
high $p_\perp$  processes is  indeed a perturbation,  unlikely to
alter  the overall  dynamics.  The  simulation  definitely treats
hard processes, if only through the observed behaviour in NN over
a range of energy.

We note that the second phase (II) is truly a much softer cascade
with  the collisions between  prehadrons, for  example, involving
energies less than ${s}^{1/2}\sim 15$ GeV and even lower averages.
These  energy  levels  in  II  were  arrived  at  by  sharing  of
energy-momentum in  the original baryon groups  with the produced
mesonic prehadrons.  Since much  of the physics here results from
stage II of the simulation  it is permissible to use soft physics
as a driving force in the overall dynamics.
 
That  the  high $p_\perp$  prehadrons  produced  should be  given
hadronic-like     cross-sections     perhaps    requires     some
justification. We  offer two possibilities.  Molnar~\cite{molnar}
has tried to correct the 'flow' deficiencies of partonic cascades
by  introducing coalescence  of  quark-like partons  at an  early
stage  in the  cascade.  Clearly  such  coalescence probabilities
will decrease steeply with increasing transverse momentum, at say
small rapidity,  but that is just  what is observed  in the meson
production  from  both  PP  and  AA.   We  then  argue  that  the
coalescence   probability  will  be   greatest  for   the  larger
transverse-sized mesons  produced and such  objects will dominate
the dynamics.

Furthermore,  and actually relevant  to our  specific simulation,
the earliest  meson-meson collisions in our  second stage cascade
II have a  first peak in time at some .25-.35  fm/c while in fact
collisions extend  to considerably later time.  This permits even
smaller prehadrons to have appreciably increased their transverse
size before collision and simultaneously suggests most collisions
are between  comovers.  Also,  the premesons, which  dominate the
second   stage  dynamics,   are  given   only  $4/9$   the  total
cross-section  of  baryon-baryon  and  hence  need  only  possess
considerably reduced effective tranverse sizes, than say baryons.

\subsection{Initial Conditions for II}

The final operation  in stage I is to  set the initial conditions
for the hadronic cascade  in stage II.  The energy-momentum taken
from the initial baryons and shared among the produced prehadrons
is  established  and an  upper  limit  placed  on the  production
multiplicity   of  prehadrons  and   normal  hadrons.    A  final
accounting of  energy sharing is  carried out through  an overall
4-momentum conservation  requirement.  We emphasize  that this is
carried out separately within each group of interacting nucleons.

The spatial  positioning of the  particles at this time  could be
accomplished in a  variety of ways.  We have  chosen to place the
prehadrons  from  each  group   inside  a  cylinder,  for  a  A+B
collision, given the initial longitudinal size of the interaction
region at each  impact parameter.  We then allow  the cylinder to
evolve   freely   according    to   the   longitudinal   momentum
distributions,  for a fixed  time $\tau_p$,  defined in  the rest
frame  of  each  group.   At   the  end  of  $\tau_p$  the  total
multiplicity  of the  prehadrons  is limited  so  that, if  given
normal hadronic  sizes appropriate to  meson-meson cross-sections
$\sim (2/3)(4\pi/3) (0.6)^3$ fm$^{3}$, they do not overlap within
the cylinder.  Such a limitation in density is consonant with the
general  notion  that  produced   hadrons  can  only  exist  when
separated  from   the  interaction  region  in   which  they  are
generated~\cite{gottfried}.  One can  conclude from this that the
prehadron  matter  acts like  an  incompressible  fluid, viz.   a
liquid,  a  state described  in  the  earliest calculations  with
LUCIFER~\cite{luc4brahms,lucifer1}.

Up  to this  point  longitudinal boost  invariance is  completely
preserved since stage I is carried out using straight line paths.
The technique  of defining the  evolution time in the  group rest
frame is essential to  minimizing residual frame dependence which
inevitably  arises in  any  cascade, hadronic  or partonic,  when
transverse momentum  is present. This  due to the finite  size of
the  colliding  objects  implied  by their  non-zero  interaction
cross-section.  

The   collision  history  recorded   enumerates  the   number  of
interactions suffered by any baryon group member and allows us to
assign a transverse momentum through random walk.  The premesonic
multiplicities,  subject to the  maximum density  injunction, are
obtained using  the KNO scaling  we invoke for the  elementary NN
interactions and the known  dependence on flavour. The production
of  baryon-antibaryon  pairs  is  allowed,  guided  again  by  NN
constraints.  Transverse momentum is generated for the prehadrons
and other produced particles, paying attention to the random walk
increase for mesons created by multiple baryon-baryon collisions.
Finally, overall energy-momentum conservation is imposed and with
it the multi-particle phase space defined.

\section{Stage II: Final State Cascade}

Stage  II is  as stated  a straightforward  cascade in  which the
prehadronic  resonances  interact  and  decay as  do  any  normal
hadrons  present or  produced during  this  cascade.  Appreciable
energy having being finally transferred to the produced particles
these  `final  state' interactions  occur  at considerably  lower
energy than  the initial  nucleon-nucleon collisions of  stage I.
This  cascade of course  imposes energy-momentum  conservation at
the two body collisional level and leads to additional transverse
momentum    production.     For     Au+Au,    the    effect    of
prehadron-prehadron  interaction  is  truly appreciable,  greatly
increasing multiplicities and  total transverse energy, $E_\perp$
through both production and  eventual decay into the stable meson
species. But  also, as  we will see,  in the suppression  of high
$p_\perp$ production.

We are  then in a position  to present results for  Au+Au 200 GeV
collisions.   These  appear  in  Figs(3--7)  for  various  double
differential  tranverse  momenta   spectra  or  their  derivative
ratios.         Most        contain       comparisons        with
PHENIX~\cite{phenix,phenixpp}  $\pi^0$  measurements.  In  future
work  we consider also  charged data,  but there  produced proton
spectra  play   an  increasingly  larger   role  with  increasing
$p_\perp$. One expects similar global behaviour.

The initial  conditions created to start the  final cascade could
have perhaps  been arrived  at through more  traditional, perhaps
partonic, means.  The second stage would then still proceed as it
does here.  We reiterate that  our purpose has been to understand
to what extent the results  seen in Figures (1-7) are affected by
stages  I and  II  separately.  {\it  i.~e.}~do  they arise  from
initial or from final state interactions.

\section{Results: Comparison with $\pi^0$ Data}

Figs. (1)  and (2)  speak to the  input nucleon-nucleon  data and
this   elementary    production   of   $\pi^0$'s    and   charged
particles. These transverse momentum spectra at mid-rapidity have
been   compared  to   results  from   PHENIX~\cite{phenixpp}  and
UA1~\cite{ua1}.  Fig(3) contains the simulated $\pi^0$ transverse
momenta  spectrum  for Au+Au  at  $\eta=0$  alongside the  PHENIX
data~\cite{phenix}.  To  a large extent  the suppression observed
experimentally is paralleled  by the simulated calculations.  The
production time  $\tau_p$ introduced  above was given  two values
$(2R_{Au}/\Gamma_{average})$ and twice this value.  The variation
with this  initial state time,  a parameter in our  modeling, was
small.  The  $\Gamma$'s here are the  longitudinal Lorentz factor
defined above and introduced  for each baryon group separately in
its rest frame.  For the symmetric Au+Au collision this change of
time  is   only  of   small  consequence  for   minimising  frame
dependence.

What conclusions are to be  drawn from these first results?  Most
definitely, one  cannot ignore final  state cascading.  Moreover,
for  the assumptions  we have  made, the  most crucial  being the
perhaps  early commencement  of such  cascading,  the suppression
cannot be  considered as necessarily  a sign for production  of a
quark-gluon  plasma: perhaps  only a  prehadron  dominated medium
after  some initial  delay.   We repeat:  despite the  apparently
short  time $\tau_p$  at say  $\sqrt{s}$ =200  GeV, one  finds in
practice  that  the second  stage  collisions  have an  effective
production  or delay time  of $0.25-0.35$  fm/c and  continue for
some tens of fermis/c. Thus `comoving' collisions likely dominate
the   cascade.   Since   the  pQCD   approach  is   clearly  more
fundamental, provided a clear  treatment of soft processes can be
included, one cannot  rule it out as a  basis for interpretation.
But  surely it is  interesting to  pursue an  alternative, albeit
more  phenomenological,  nuclear-system  oriented view.   A  view
which simply  suggests that more  detailed and definite  signs of
QCD plasma must be sought,  especially from the earliest times in
the ion-ion collision.

It is instructive to  deconstruct the elements of the simulation,
~i.e. to separate  the spectrum at the conclusion  of I from that
resulting  from both  I+II.   In Fig.(4)  the $\pi^0$  transverse
momentum  yield  is  shown  for  both  these  cases  against  the
experimental  data.   It is  immediately  evident  that the  many
virtual  NN  collisions  in  stage  I  produce  a  much  elevated
$p_\perp$ output and that this is in turn reduced by more than an
order of magnitude by collisions with presumably other prehadrons
in II. Part  of this effect is through  inclusion in the dynamics
of at  least a  kinematical treatment of  energy loss.   Thus the
above  analogy  of an  initial  hot  gas  cooled by  final  state
interactions during expansion, is apt.

One might well turn this  around and declare that the final state
scattering of  a given prehadron  with comovers has cut  down the
Cronin effect,  a reduction  which suggests the  applicability of
the term `jet suppression'  through final state interactions. One
notes parenthetically  that particles lost at  high $p_\perp$ are
compensated for by an increase at the lowest $p_\perp$'s. This is
of course part of the effective cooling observed.

A second  and equally important  criterion for the  simulation is
the  maximum  densities created  as  initial  conditions for  II.
Fig(5) cast some  light on this and on  another issue, the actual
transverse energy  density attributed  to the earliest  stages of
the  collision.  We  have  included in  this  figure the  charged
$dN/d\eta$  spectra,  including  a  BRAHMS charged  meson  result
(compared to a simple fit):

(a) for the totality of `stable' mesons in I+II,

(b) the  same result for I  alone when only  decays of prehadrons
but no stage II reinteractions are permitted,

(c) and finally for the prehadrons in I only, with no decays.

It  is evident  that some  $2/3$ of  the summed  tranverse energy
$E_\perp$ is generated in the  second expanding phase II when the
system is  increasing both longitudinally  and transversely.  The
initial,  very   early,  $E_\perp$  generated   is  then  reduced
commensurably, falls  well below the  Bjorken limit and  is hence
not all  available for  initial `plasma' generation.   In present
calculations at the initialisation of II, and keeping in mind the
average masses assigned to  prehadrons centered at $0.8$ to $1.0$
GeV,  the  ambient  transverse  energy densities  are  $\le  1.8$
GeV/$fm^3$)  for the  shortest initial  time $\tau_p$  chosen and
correspondingly less for longer times.

It is also clear that the density of prehadrons, each of which in
I as seen  from Fig.(5) decays into some  2.5 stable hadrons. The
rather lengthy decay time assigned, $\tau_f \sim 1$ fm/c, ensures
that the stable  species enter only sparingly in  the dynamics of
II.   Thus the  bugbear of  over-scattering or  over-cascading is
reduced to manageable proportion. The initial limit on density is
of course an additional key element.
 
\subsection{Ratios of Au+Au to NN Production}

In  Fig.(6) we  display the  oft-quoted ratio~\cite{phenix4,star,
brahms,phobos} introduced in  Eqn.(1).  The discrepancies between
simulation  and data  are  magnified but  the  general effect  of
agreement  shown   in  Fig.(3)   is  left  intact.    The  direct
double-differential cross-sections fall  many orders of magnitude
over   the  reported   range   but  the   ratios  to   normalized
proton-proton  cover a  much compressed  range over  the reported
$p_\perp$ interval.   We have in this figure  presented PHENIX $\
pi^0$  data from  an early  paper~\cite{phenix}, but  more recent
data~\cite{shimamura} tells a somewhat  modified tale. The dip in
${R}_{AuAu/NN}$   at  the   lowest   transverse  momentum   point
$p_\perp$= 1.25 GeV/c, implying a  maximum in this ratio at a bit
higher  $p_\perp$, has  apparently disappeared.   This is  so for
both the  $0-5\%$ and $0-10\%$  data resulting in curves  more in
accord  with  the  corresponding theoretical  calculations.   The
equivalent  charged  ratios~\cite{brahms,phenix4,phobos,star}  do
exhibit this  apparent maximum but  in a region where  the proton
spectra show  considerable activity. One might  also question the
use of the ratio $R_{AA/NN}$  at very low $p_\perp$, involving as
it does the number of  binary collisions. A relevant or preferred
divisor there might be the number of participants.

In Fig.(6) we present  LUCIFER results for $0\%-10\%$ centrality,
but  the  shape  is  generic  at  least  for  reasonably  central
collisions.   As one approaches  extreme peripherality  the ratio
will  at all  $p_\perp$  eventually approach  unity, provided  of
course  the  number  of  binary  collisions  $N_{coll}$  is  also
adjusted.

It must  be emphasized that  $R_{AuAu/NN}$ is quite  sensitive to
the  invariant   $PP$  cross-section.   We   have  prepared  both
theoretical and experimental ratios using the PHENIX experimental
$PP$ values, indeed using our  own constructed fits to the latter
data and  for smoothness  to the simulated  $Au+Au$.  The  fit to
PHENIX $PP$ is  shown in Fig.{7}.  It should  be again noted that
the prehadron-liquid  modeling employed here is  to be considered
somewhat metaphorically  and that no  additional parameters, over
and above  the times for production,  $\tau_p$, and hadronisation
(decay  of   prehadrons),  $\tau_f$,  were   made  available  for
sculpting the  data, nor  would such editorializing  be sensible.
We   seek  only   a  qualitative   and   reasonably  quantitative
understanding of the observations.

The simulations performed here overestimate, but not by much, the
suppression of the spectrum at the lowest $p_\perp$, but yield an
overall, perhaps surprisingly, accurate description.

\section{Conclusions}

Repeating: an alternative reading of the high transverse momentum
suppression  has   been  presented.   It  is   hard  to  conclude
definitively from what is presented here that the standard pQCD +
some  soft process  treatment  is not  still  a more  fundamental
approach.   But  at  question  is  just  this  coupling  to  soft
processes,  not  unrelated to  possible  early  appearance of  an
excitation spectrum of  hadronic-like structures.  The latter, if
generated sufficiently early may alter the role of soft processes
even on high $p_\perp$ objects  passing through a much more dense
cloud of soft prehadronic. It is fact just the nature of the soft
processes  which is  at  question.  Certainly  there  is still  a
silent  elephant  lurking in  the  dynamics:  the observation  of
rather large  elliptical flow in  the meson spectrum~\cite{flow}.
These flow measurements  are most easily reproduced theoretically
if hadronic-sized coss-sections are invoked~\cite{molnar}.

In  the  work  on   D+Au~\cite{luc4brahms}  the  use  of  such  a
prehadronic  spectrum exposed  most  clearly the  simple role  of
dynamically-driven geometry in ratios of BRAHMS $p_\perp$ spectra
at   varying  pseudorapidity.   In   D+Au  high   $\eta$  charged
production,   with  attendant   higher   energies  for   produced
particles,  results mostly from  nucleon-nucleon collisions  in a
ring around the  periphery of the target, at  a reduced frequency
in comparison to $\eta$=  0, essentially a volume process.  Hence
the  $\sim  2$   suppression  seen  in  BRAHMS~\cite{brahms}  for
$\eta=3.2$ yield relative to $\eta=0$.  For Au+Au the degradation
in  $p\perp$ spectra  in  the phase  II  cascade is  considerably
increased and suppression, when appropriately to NN, is seen even
for  $\eta=0$.   Certainly,  the  RHIC  experiments  are  probing
unusual nuclear  matter, at high hadronic and  energy density, in
exciting terms and more experimental exploration is required.

It  would  seem,  however,  that  a  direct  attempt  at  a  pQCD
explanation of this behaviour must claim that, at the very least,
all soft mesons are produced in essentially hard collisions.  The
presentation here provides an interesting case for relying on the
geometry of soft,  low $p_\perp$, processes, essentially mirrored
in hard processes, to produce  the major features of the D+Au and
Au+Au   data.   True   enough,  the   high  $p_\perp$   tails  in
distributions  are   merely  tacked  on  in   our  approach,  but
legitimately  so   by  using  the   NN  data  as  input   to  the
nucleus-nucleus cascade.  In any  case one should again very much
be  cognizant of  the small  number of  high  $p_\perp$ particles
present in  even cental collisions at the  highest RHIC energies.
Some $5\%-10\%$  of the integrated spectrum of  mesons comes from
$p_\perp \ge 1$ GeV in an average NN event, and similar levels in
$Au+Au$.

\section{Acknowledgements}
This  manuscript  has  been  authored  under  the  US  DOE  grant
NO. DE-AC02-98CH10886. One of  the authors (SHK) is also grateful
to the  Alexander von Humboldt Foundation, Bonn,  Germany and the
Max-Planck   Institute  for   Nuclear  Physics,   Heidelberg  for
continued  support and hospitality.   Useful discussion  with the
BRAHMS,  PHENIX, PHOBOS  and STAR  collaborations  are gratefully
acknowledged, especially  with C.~Chasman, R.~Debbe, F.~Videbaek.
D.~Morrison, M.~T.~Tannenbaum, T.~Ulrich and J.~Dunlop.

\vfill\eject

\begin{figure}
\vbox{\hbox to\hsize{\hfil
\epsfxsize=6.1truein\epsffile[0 0 561 751]{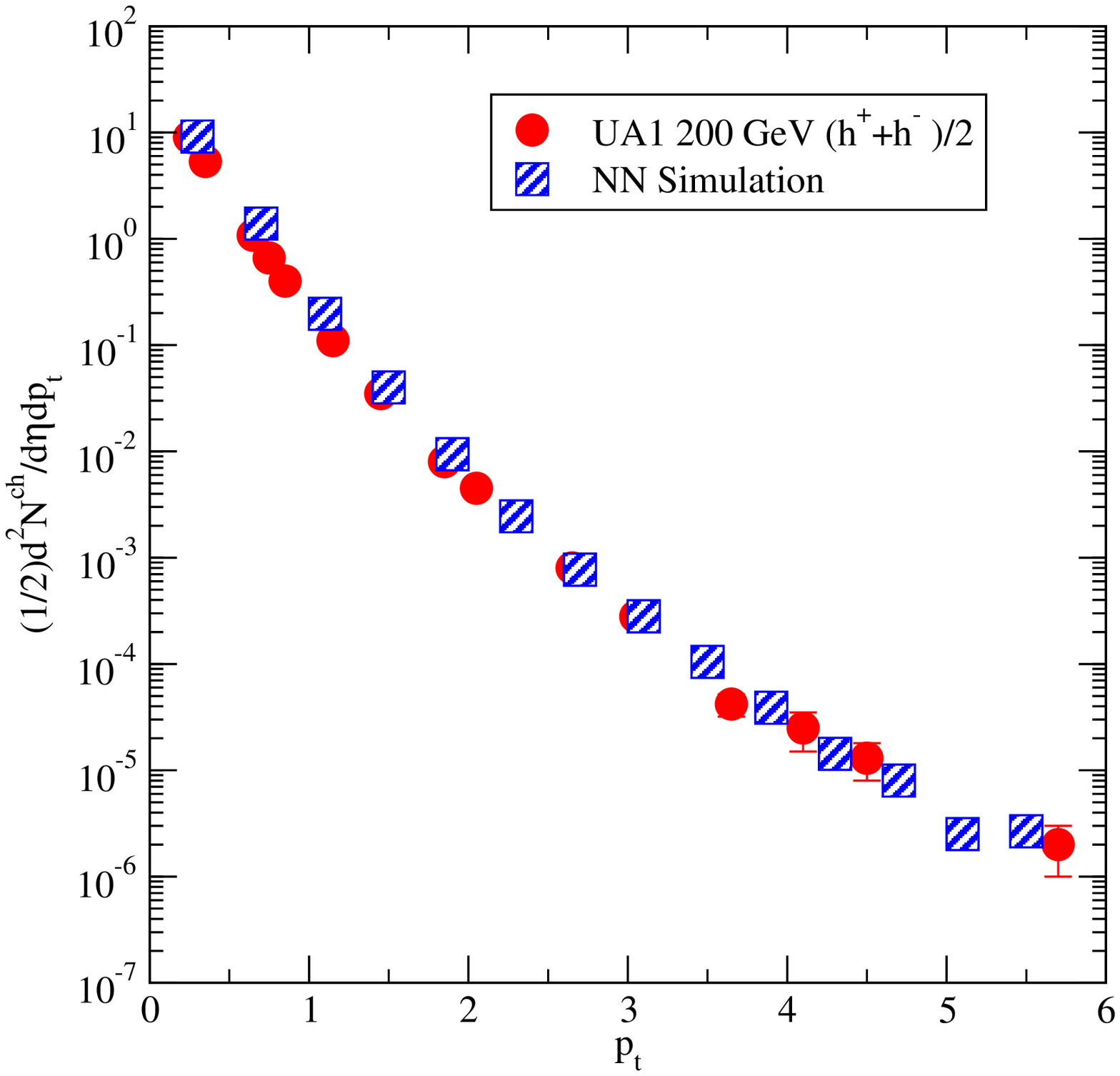}
\hfil}}
\caption[]{PP Pseudorapidity  spectra: Comparison of  UA1 minimum
bias  200 GeV  NSD  data~\cite{ua1} with  an appropriate  LUCIFER
simulation. The  latter is properly obtained  from experiment and
an input to the ensuing AA collisions; thus does not constitute a
`set' of free parameters.}
\label{fig:Fig.(1)}
\end{figure}
\clearpage

\begin{figure}
\vbox{\hbox to\hsize{\hfil
\epsfxsize=6.1truein\epsffile[0 0 561 751]{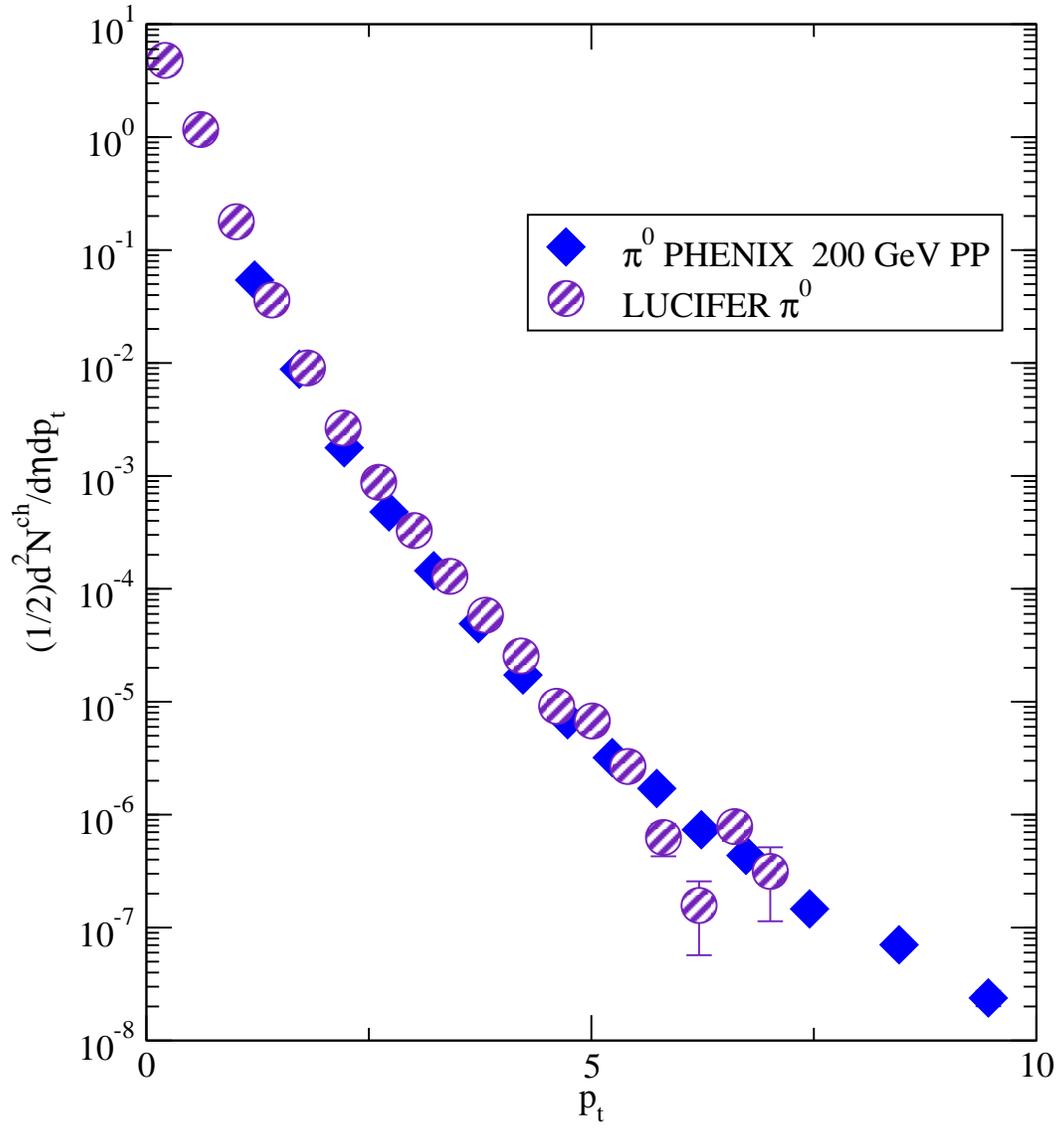}
\hfil}}
\caption[]{A  similar transverse  momentum  $\pi^0$ spectrum  from
PHENIX PP~\cite{phenixpp} vs simulation.}
\label{fig:Fig.(2)}
\end{figure}
\clearpage

\begin{figure}
\vbox{\hbox to\hsize{\hfil  
\epsfxsize=6.1truein\epsffile[0 0 561 751]{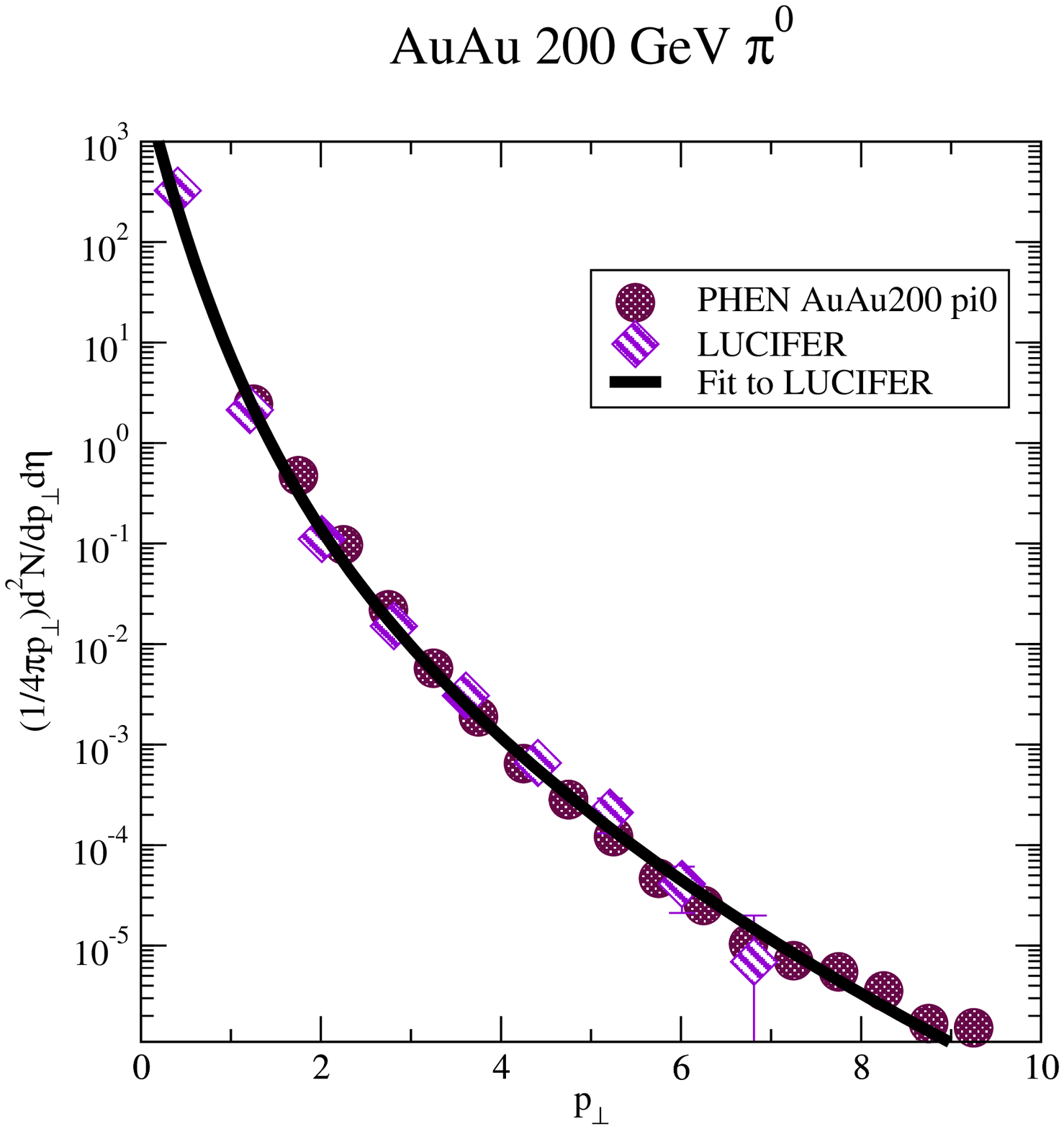} 
\hfil}}
\caption[]{Central   PHENIX  $\pi^0$   200  GeV   for   Au+Au  vs
simulation.  Curves for different  choices of the production time
$\tau_p$, say 0.02  fm/c and 0.04 fm/c differ  very little, since
in effect  the cascade  effectively begins somewhat  later, neare
$0.25-0.35$   fm/c  and   continues  much   longer  to   tens  of
fm/c.  Centrality  for PHENIX  is  here  $0\%-10\%$, roughly  for
impact parameters $b<4.25$ fm. in the simulation.}
\label{fig:Fig.(3)}
\end{figure}
\clearpage

\begin{figure}
\vbox{\hbox to\hsize{\hfil
\epsfxsize=6.1truein\epsffile[0 0 561 751]{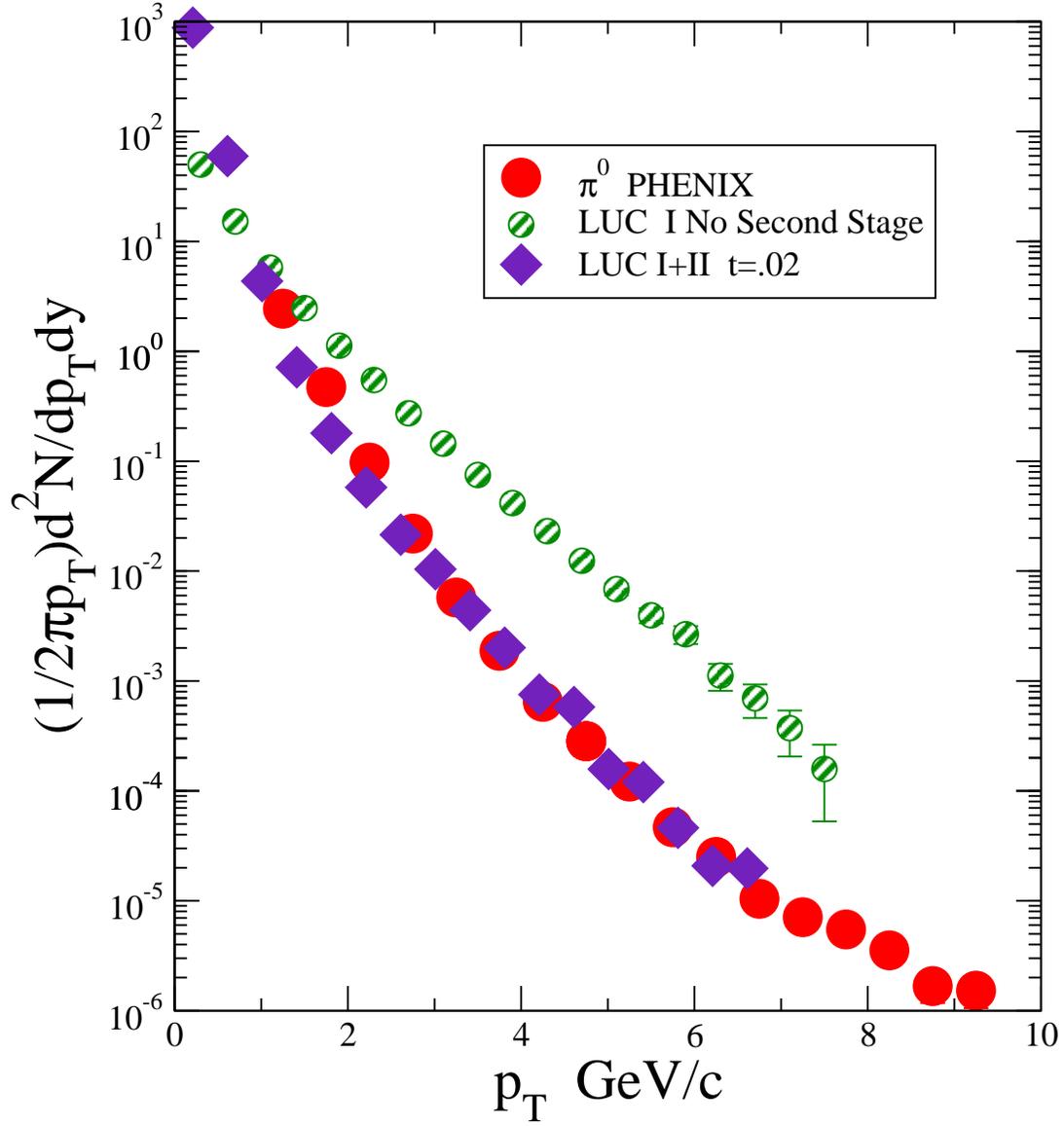}
\hfil}}
\caption[]{The $\pi^0$ transverse momenta  yields for stage I, no
final  cascade, vs  those for  the full  stage  I+II calculation.
Clearly there  is considerable suppression in  the final cascade.
Recalling that the  experimentalists quote a `direct' suppression
0f $\sim$ 4--5 for the ratio in Eqn.(1) at the highest $p_\perp$,
there is  at the end  of I an  enhancement $\sim $3,  {\it i.~e.}
still a Cronin effect in this first stage.}
\label{fig:Fig.(4)}
\end{figure}
\clearpage

\begin{figure}
\vbox{\hbox to\hsize{\hfil
\epsfxsize=6.1truein\epsffile[0 0 561 751]{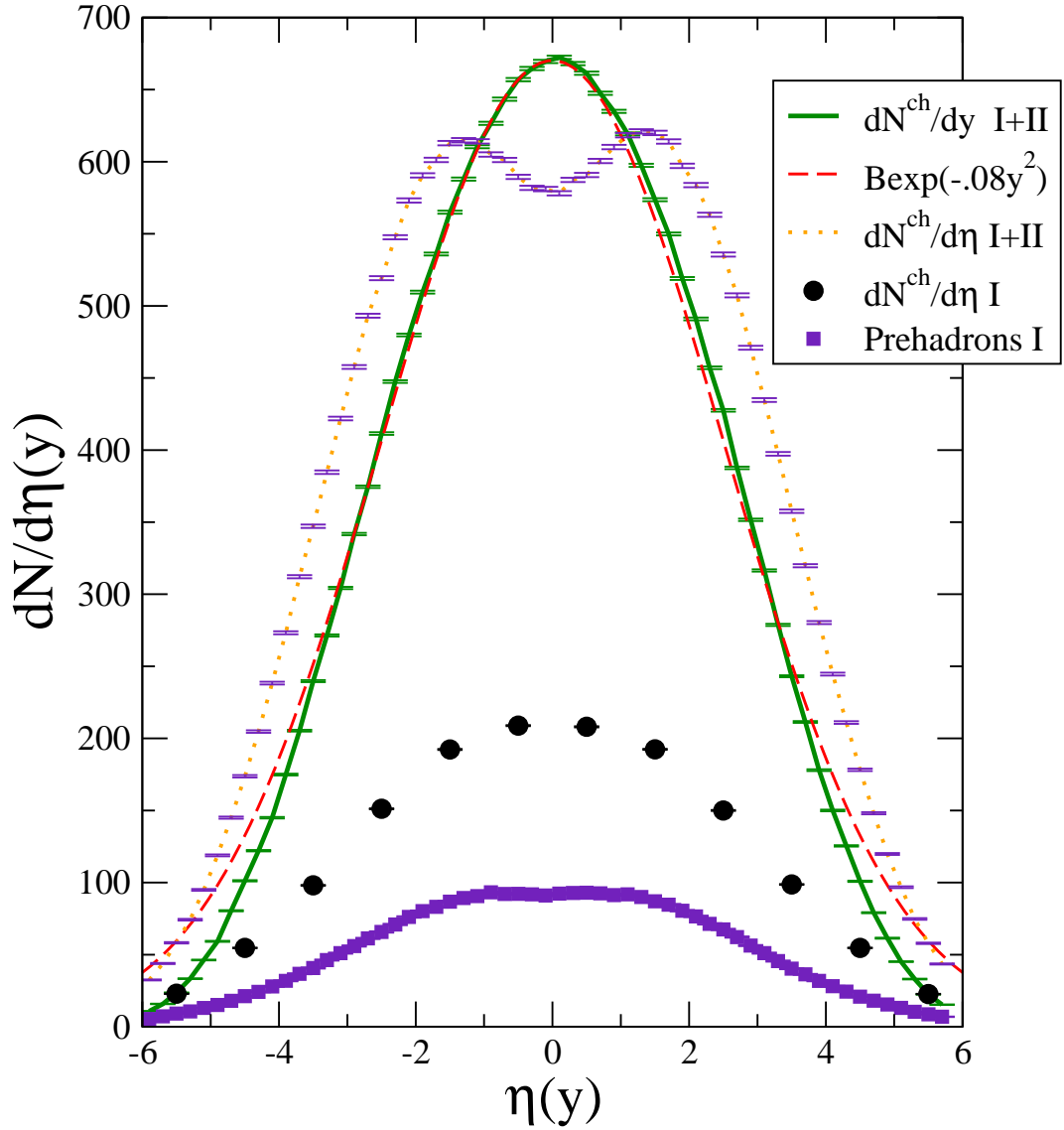}
\hfil}}
\caption[]{Pseudorapidity and rapidity spectra for charged mesons
and  prehadrons at  various stages  of the  collision simulation.
The  gaussian  fit  to  the charged  pion  rapidity  distribution
approximates  preliminary BRAHMS~\cite{brahmsfwhmprelim} results,
certainly in its FWHM, and hence demonstrates the validity of the
simulation  (solid   line)  for  $dN^{ch}/dy$.    The  successive
retreats, first  to phase  I of the  simulation and then  to only
prehadrons in phase I, {\it i.~e.}  no decays, indicates both the
reduction  in  cascading participators  and  in  the fraction  of
$E_\perp$ avalable at the earliest momemts of the cacscade.}
\label{fig:Fig.(5)}
\end{figure}
\clearpage

\begin{figure}
\vbox{\hbox to\hsize{\hfil
\epsfxsize=6.1truein\epsffile[0 0 561 751]{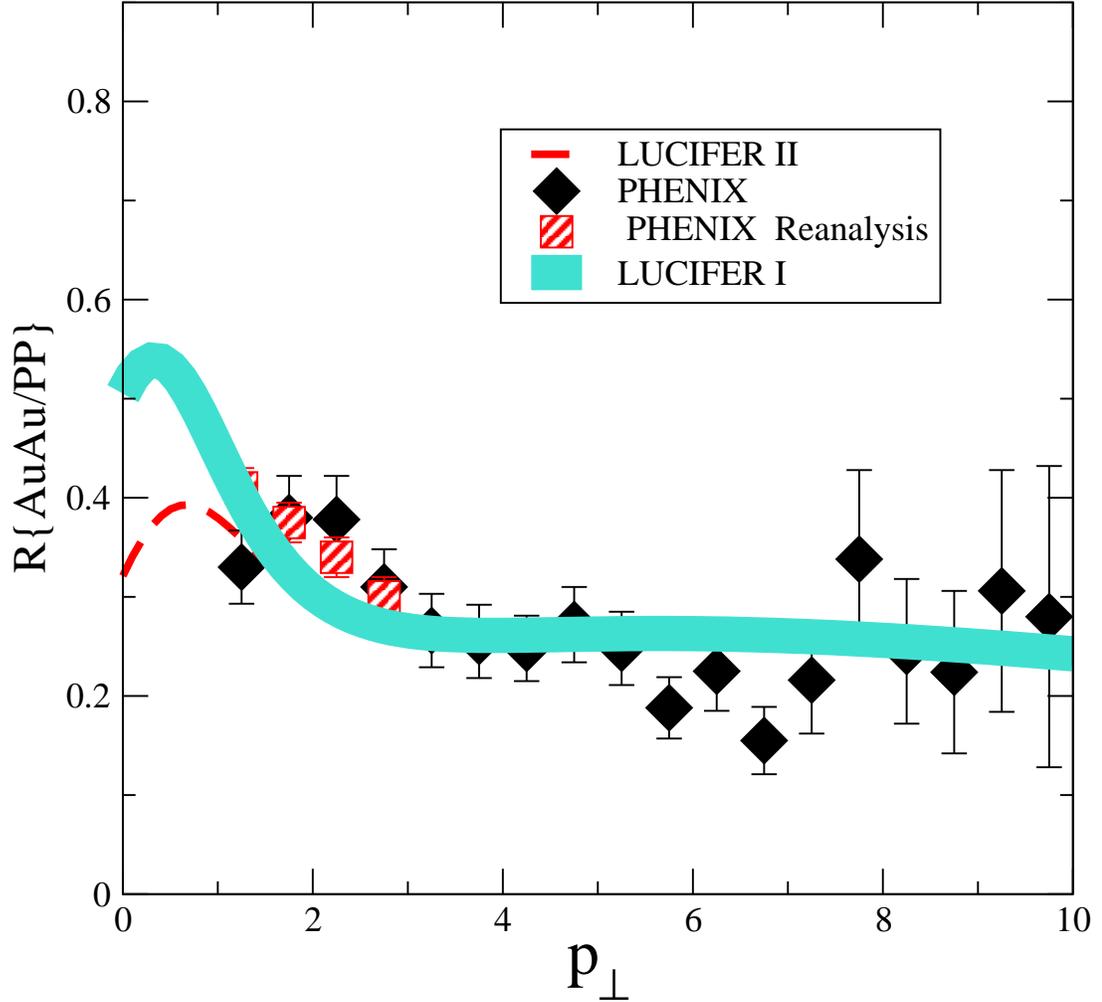}
\hfil}}
\caption[]{${R}_{AuAu/NN}$:   Ratio  of  $\pi^0$   production  in
$Au+Au$  collisions to  that in  NN  at $\sqrt{s}$=  200 GeV  for
$0\%-10\%$  centrality. The binary  collision number  selected at
this  centrality is  N$_{coll}$=950, while  the divisor  for both
PHENIX and LUCIFER curves shown here  is one fit to the PHENIX PP
data  shown  in  Fig.(7).   The  resultant  ratio  is  especially
sensitive to the PHENIX NN data. LUCIFER curves are shown for two
choices differing not in the $Au+Au$ calculations but only in the
fitted PHENIX extrapolation to  lowest $p_\perp$, where of course
there is no  data.  The extra points for the  PHENIX ratio at the
four    lowest   $p_\perp$   GeV/c    are   from    more   recent
analysis~\cite{shimamura}. Thus  the general shape  and magnitude
of  the  simulation conforms  well  with  this  most recent  data
analysis, both exhibiting a rise towards smaller $p_\perp$.  This
feature   is  also   present  in   the  very   similar  $0\%-5\%$
measurement~\cite{shimamura}.   The   calculated  suppression  is
somewhat excessive at low $p_\perp$.}
\label{fig:Fig.(6)}
\end{figure}
\clearpage

\begin{figure}
\vbox{\hbox to\hsize{\hfil
\epsfxsize=6.1truein\epsffile[0 0 561 751]{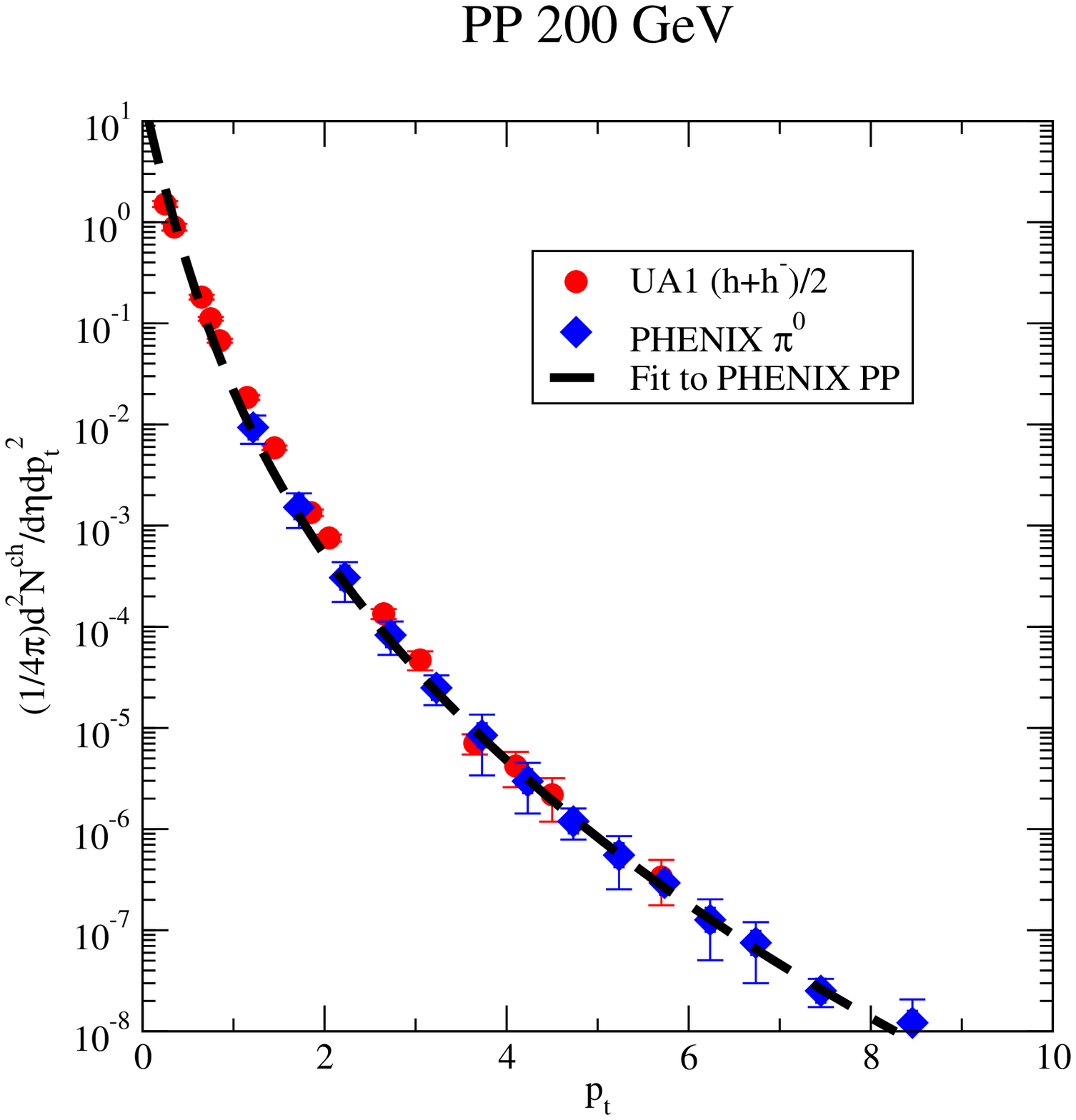}
\hfil}}
\caption[]{PHENIX $\pi^0$ 200 GeV/c for NN: The fit to the PHENIX
PP data used in the ratios  shown in Fig.(6) is shown against the
observations.}
\label{fig:Fig.(7)}
\end{figure}
\clearpage

\end{document}